\documentclass[twocolumn,showpacs,showkeys,reprint,amsmath,amssymb,aps]{revtex4-1}

\usepackage{graphicx}
\usepackage{dcolumn}
\usepackage{bm}
\usepackage{tabularx}
\usepackage{amsmath}

\begin{document}

\title{Multigap nodeless superconductivity in CsCa$_2$Fe$_4$As$_4$F$_2$ probed by heat transport}

\author{Y. Y. Huang,$^1$ Z. C. Wang,$^2$ Y. J. Yu,$^1$ J. M. Ni,$^1$ Q. Li,$^1$ E. J. Cheng,$^1$ G. H. Cao,$^{2,3}$ and S. Y. Li$^{1,3,*}$}

\affiliation
{$^1$State Key Laboratory of Surface Physics, Department of Physics, and Laboratory of Advanced Materials, Fudan University, Shanghai 200438, China\\
 $^2$Department of Physics and State Key Lab of Silicon Materials, Zhejiang University, Hangzhou 310027, China\\
 $^3$Collaborative Innovation Center of Advanced Microstructures, Nanjing 210093, China
}

\date{\today}

\begin{abstract}
Recently, a new family of iron-based superconductors called 12442 was discovered and the muon spin relaxation ($\mu$SR) measurements on KCa$_2$Fe$_4$As$_4$F$_2$ and CsCa$_2$Fe$_4$As$_4$F$_2$ polycrystals, two members of the family, indicated that both have a nodal superconducting gap structure with $s+d$ pairing symmetry. Here we report the ultralow-temperature thermal conductivity measurements on CsCa$_2$Fe$_4$As$_4$F$_2$ single crystals ($T_c$ = 29.3 K). A negligible residual linear term $\kappa_0/T$ in zero field and the field dependence of $\kappa_0/T$ suggest multiple nodeless superconducting gaps in CsCa$_2$Fe$_4$As$_4$F$_2$. This gap structure is similar to CaKFe$_4$As$_4$ and moderately doped Ba$_{1-x}$K$_x$Fe$_2$As$_2$, but contrasts to the nodal gap structure indicated by the $\mu$SR measurements on CsCa$_2$Fe$_4$As$_4$F$_2$ polycrystals.
\end{abstract}

\maketitle

Since the discovery of superconductivity in layered iron arsenide La(O$_{1-x}$F$_x$)FeAs \cite{ironfirst}, many iron-based superconductors have come to the fore \cite{122dis,111dis,11dis,21311dis1,21311dis2,1144first,12442K}. Up to now, iron-based superconductors have already developed into a diverse group of different structural families \cite{ironreview}, such as the so-called 1111, 122, 111, and 11 families \cite{ironfirst,122dis,111dis,11dis}. Meanwhile, great efforts have also been made to understand their various unconventional superconducting properties, especially, their superconducting gap structures and pairing symmetry \cite{ironreview,XHChenreview}.

Of particular interest is the 122 family, in which the parent compound $Ae$Fe$_2$As$_2$ ($Ae$ = Ba, Ca, Sr) can be turned into a superconductor through either hole or electron doping \cite{elecdoping,122dis}. For the typical hole-doped Ba$_{1-x}$K$_x$Fe$_2$As$_2$ with $x \leq$ 0.55, multiple full superconducting gaps have been demonstrated by many experimental studies \cite{gaparpes1,gaparpes2,gapdepth1,gapdepth2,gapnmr1,gapnmr2,luoBaKFeAs,gapandreev}. In contrast, the extremely hole-doped KFe$_2$As$_2$ ($x$ = 1) displays clear nodal superconducting gap \cite{KFeAsnmr,KFeAsthermal1,KFeAsdepth,KFeAsthermal2}. Such a change of the gap structure is attributed to the evolution of Fermi surface with K doping \cite{Thomale2011}, which has been revealed by angle-resolved photoemission spectroscopy (ARPES) measurements \cite{KFeAsarpes}. While the multiple nodeless gaps in the moderately doped regime of Ba$_{1-x}$K$_x$Fe$_2$As$_2$ are consistent with the $s_{\pm}$-wave pairing \cite{Thomale2011,gapnmr1,gapandreev}, it is still under debate whether the nodal gap in the heavily overdoped regime represents a fundamental change of the pairing symmetry. Some suggested that further doping causes the pairing symmetry to change into the $d$-wave pairing \cite{Thomale2011,KFeAsdepth,KFeAsthermal2}, however others argued accidental gap nodes in heavily overdoped Ba$_{1-x}$K$_x$Fe$_2$As$_2$ \cite{KFeAsswave,XCHongCPL}. Note that RbFe$_2$As$_2$ and CsFe$_2$As$_2$ also inherit the nodal superconducting gap structure of KFe$_2$As$_2$ \cite{CsFeAs,RbFeAs}.

In order to avoid the substitutional disorder in Ba$_{1-x}$K$_x$Fe$_2$As$_2$, stoichiometric CaKFe$_4$As$_4$ (1144) was first synthesized in 2016 \cite{1144first}. Unlike the random occupation of Ba$^{2+}$ and K$^{+}$ ions in Ba$_{1-x}$K$_x$Fe$_2$As$_2$, the Ca$^{2+}$ and K$^{+}$ ions in CaKFe$_4$As$_4$ form alternating planes along the crystallographic $c$ axis, separated by FeAs layers \cite{1144first}. Soon afterwards, a variety of studies revealed the multiple nodeless superconducting gaps and $s_{\pm}$-wave pairing symmetry of CaKFe$_4$As$_4$ \cite{1144properties,1144arpes,1144nmr,1144musr,1144depth1,1144depth2,1144ins}, pretty like the moderately doped Ba$_{1-x}$K$_x$Fe$_2$As$_2$.

Recently, a new family of iron-based superconductor called 12442 was designed by replacement of the Ca layers in $A$CaFe$_4$As$_4$ ($A$ = K, Rb, Cs) with Ca$_2$F$_2$ layers and synthesized successfully \cite{12442K,growth}. A series of $\mu$SR measurements on the polycrystalline 12442 compounds suggested the presence of line nodes in the superconducting gaps of KCa$_2$Fe$_4$As$_4$F$_2$ and CsCa$_2$Fe$_4$As$_4$F$_2$ \cite{musr12442K,musr12442Cs}. Since the replacement of Ca layers by Ca$_2$F$_2$ layers does not introduce additional carriers, the 12442 compounds should have the same doping level as in 1144 compounds \cite{12442K}. Thus, the nodal superconducting gap structure of 12442 compounds evidenced by $\mu$SR measurements is very striking, which motivates us to investigate their gap structure by measuring the ultralow-temperature thermal conductivity of 12442 single crystals.

In this Rapid Communication, we grew and characterized CsCa$_2$Fe$_4$As$_4$F$_2$ single crystals, then measured their ultralow-temperature thermal conductivity. A negligible residual linear term $\kappa_0/T$ in zero field is confirmed by two samples and a slow field dependence of $\kappa_0/T$ at low field is also observed. These results suggest that CsCa$_2$Fe$_4$As$_4$F$_2$ has multiple nodeless superconducting gaps, just as CaKFe$_4$As$_4$ and Ba$_{0.75}$K$_{0.25}$Fe$_2$As$_2$.

High-quality plate-like single crystals of CsCa$_2$Fe$_4$As$_4$F$_2$ were grown from CsAs flux \cite{ZCWanggrow}. The X-ray diffraction (XRD) measurement was performed by an X-ray diffractometer (D8 Advance, Bruker). The DC magnetization measurement was performed down to 1.8 K using a magnetic property measurement system (MPMS, Quantum Design). The specific heat was measured in a physical property measurement system (PPMS, Quantum Design) via the relaxation method. Two samples, labeled as A and B, were cut to a rectangular shape with dimensions of $\sim$ 2 $\times$ 1 mm$^2$ in the $ab$ plane and a thickness of 50 $\mu$m along the $c$ axis for the transport measurements. The in-plane resistivity was measured from 273 to 1.5 K in a $^4$He cryostat. The in-plane thermal conductivity was measured in a dilution refrigerator by using a standard four-wire steady-state method with two RuO$_2$ chip thermometers, calibrated {\it in situ} against a reference RuO$_2$ thermometer. Magnetic fields were applied perpendicular to the $ab$ plane.

Figure 1(a) displays the crystal structure of CsCa$_2$Fe$_4$As$_4$F$_2$, whose Fe$_2$As$_2$ layers are surrounded by Cs atoms on one side and Ca$_2$F$_2$ layers on the other, leading to two distinct As sites above and below the Fe plane, similar to CaKFe$_4$As$_4$. The largest natural surface of the as-grown single crystals was determined as (001) plane by XRD, as illustrated in Fig. 1(b).

\begin{figure}
\includegraphics[clip,width=8.6cm]{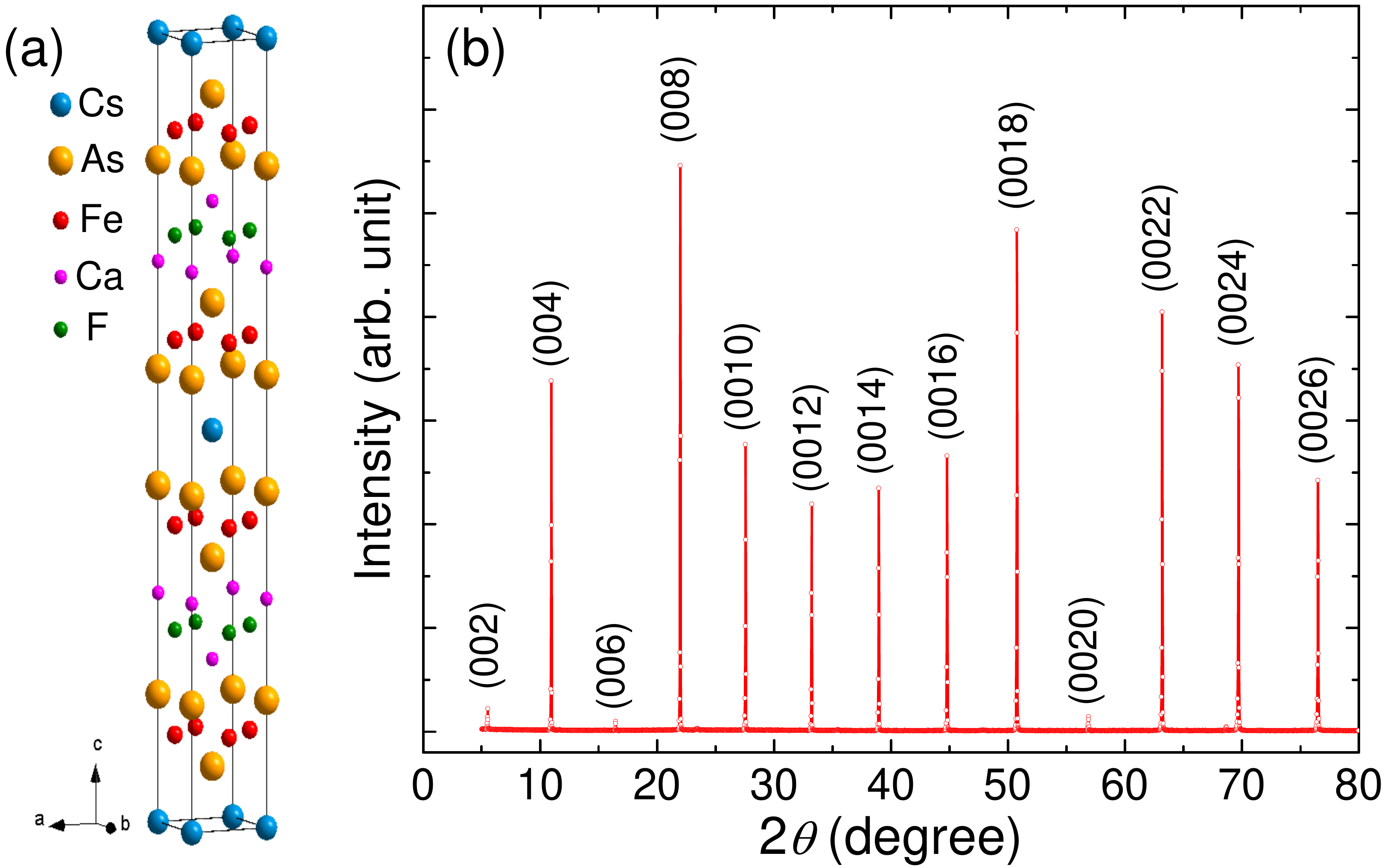}
\caption{(a) Crystal structure of CsCa$_2$Fe$_4$As$_4$F$_2$. The Cs, As, Fe, Ca and F atoms are presented as blue, orange, red, magenta and green balls, respectively. (b) Room-temperature XRD pattern from the large natural surface of the CsCa$_2$Fe$_4$As$_4$F$_2$ single crystal. Only (00$l$) Bragg peaks were found.}
\end{figure}

\begin{figure}
\includegraphics[clip,width=6.5cm]{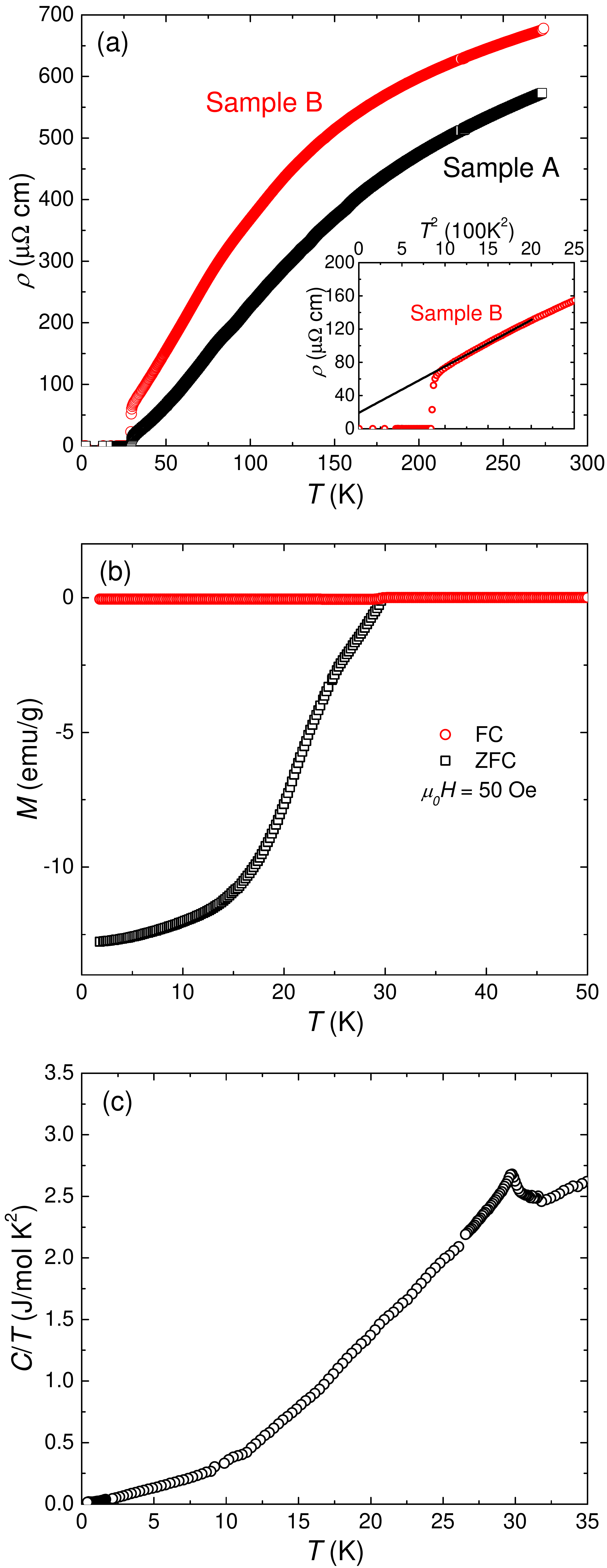}
\caption{(a) Temperature dependence of in-plane resistivity for two CsCa$_2$Fe$_4$As$_4$F$_2$ single crystals (labeled as Sample A and Sample B, respectively) in zero field. The two samples share the same critical temperature $T_c$ = 29.3 K, defined by $\rho$ = 0. Inset: same data of sample B plotted as a function of $T^2$. The solid line is a fit to $\rho$ = $\rho_0$ + $AT^2$ between 32 K and 45 K, giving the residual resistivity $\rho_0$ = 19.2 $\mu\Omega$ cm. (b) DC magnetization of CsCa$_2$Fe$_4$As$_4$F$_2$ single crystal at $\mu_0H$ = 50 Oe along $c$ axis, with zero-field and field cooling modes, respectively. The diamagnetic transition occurs at 29.2 K in both modes. (c) Temperature dependence of specific heat divided by temperature $C/T$ for the CsCa$_2$Fe$_4$As$_4$F$_2$ single crystal in zero field. Specific heat anomaly due to the superconducting transition is observed at 30.2 K.}
\end{figure}

Figure 2(a) shows the in-plane resistivity for two CsCa$_2$Fe$_4$As$_4$F$_2$ single crystals labeled as sample A and sample B down to 1.5 K in zero field. The two samples both exhibit metallic behavior without any phase transition until the sharp superconducting transition occurs. The $T_c$s of both samples are identical, 29.3 K defined by $\rho$ = 0. Previously, the polycrystalline CsCa$_2$Fe$_4$As$_4$F$_2$ exhibits similar metallic behavior with a lower $T_c$ = 28.2 K \cite{growth}. In the inset of Fig. 2(a), the $\rho(T)$ of sample B between 32 K and 45 K in the normal state can be described by Fermi liquid behavior $\rho$ = $\rho_0$ + $AT^2$, and the fit gives the residual resistivity $\rho_0$ = 19.2 $\mu\Omega$ cm. Thus the residual resistivity ratio RRR = $\rho$(273 K)/$\rho_0$ is 35.4, indicating the high quality of our single crystals.

Temperature dependence of the magnetic susceptibility from 2 to 50 K at 50 Oe in both zero-field and field cooling modes is plotted in Fig. 2(b). The diamagnetic transition occurs at 29.2 K, which is consistent with the resistivity measurement. Clear specific heat anomaly can also be seen at the superconducting transition of CsCa$_2$Fe$_4$As$_4$F$_2$ single crystal, as displayed in Fig. 2(c). This behavior of the specific heat is very similar to those of CaKFe$_4$As$_4$ and Ba$_{0.6}$K$_{0.4}$Fe$_2$As$_2$ \cite{gapheat2,1144properties}. Both susceptibility and specific heat results demonstrate the bulk superconductivity in our single crystals.

\begin{figure}
\includegraphics[clip,width=6.5cm]{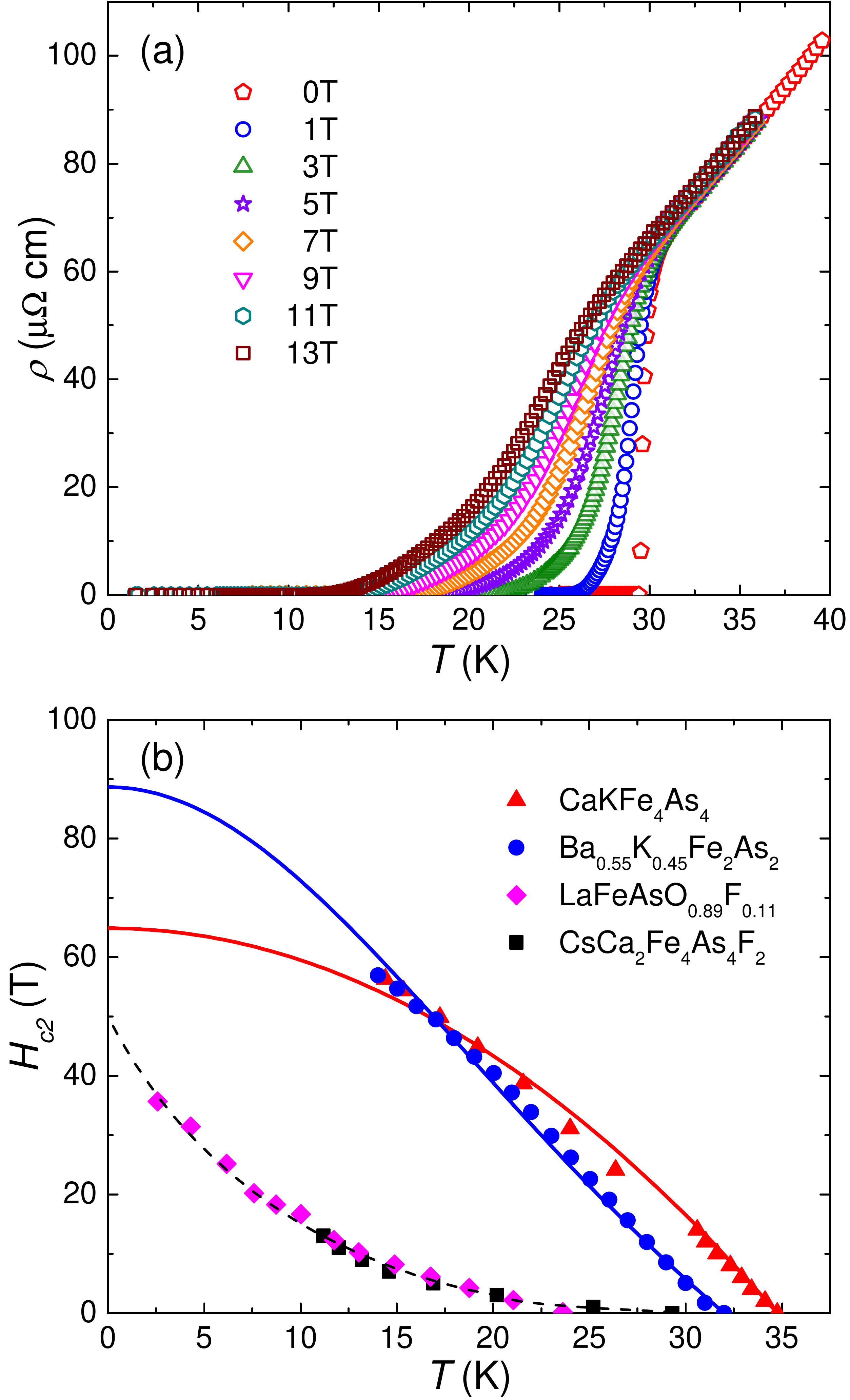}
\caption{(a) Low-temperature resistivity of CsCa$_2$Fe$_4$As$_4$F$_2$ sample B in magnetic fields up to 13 T. (b) Temperature dependence of the upper critical field $H_{c2}$($T$). For comparison, similar data of Ba$_{0.55}$K$_{0.45}$Fe$_2$As$_2$ and CaKFe$_4$As$_4$ are also shown and fitted to Ginzburg-Landau equation $H_{c2}(T)$ = $H_{c2}$(0)[1 $-$ ($T/T_c$)$^2$]/[1 + ($T/T_c$)$^2$] and empirical formula $H_{c2}$($T$) = $H_{c2}$(0)[1 $-$ ($T/T_c$)$^2$], respectively \cite{BakFeAsHc2,1144properties,GLequation}. The data of LaFeAsO$_{0.89}$F$_{0.11}$ \cite{La1111Hc2} and our CsCa$_2$Fe$_4$As$_4$F$_2$ show a concave tendency. The dashed line is a guide to the eye, from which we roughly estimate $H_{c2}$(0) $\approx$ 50 T for CsCa$_2$Fe$_4$As$_4$F$_2$.}
\end{figure}

To determine the upper critical field $H_{c2}$(0) of CsCa$_2$Fe$_4$As$_4$F$_2$, we measured the resistivity of sample B below 36 K in various magnetic fields up to 13 T, as shown in Fig. 3(a). The $H_{c2}(T)$ obtained from Fig. 3(a) is plotted in Fig. 3(b), compared with the $H_{c2}(T)$s of CaKFe$_4$As$_4$, Ba$_{0.55}$K$_{0.45}$Fe$_2$As$_2$ and LaFeAsO$_{0.89}$F$_{0.11}$ \cite{BakFeAsHc2,1144properties,La1111Hc2}. Note that the $H_{c2}(T)$ of LaFeAsO$_{0.89}$F$_{0.11}$ is defined by $\rho$ = 10\%$\rho_N$ \cite{La1111Hc2}, while the others are defined by $\rho$ = 0 \cite{BakFeAsHc2,1144properties}. The significant broadening of the superconducting transition in field and the concave $H_{c2}(T)$ are attributed to the strong two-dimensionality of CsCa$_2$Fe$_4$As$_4$F$_2$ \cite{ZCWanggrow}. Here we roughly estimate $H_{c2}$(0) $\approx$ 50 T for CsCa$_2$Fe$_4$As$_4$F$_2$, following the trend of the high-field data of LaFeAsO$_{0.89}$F$_{0.11}$. Note that a slightly different $H_{c2}$(0) does not affect our discussion on the field dependence of $\kappa_0/T$ below.

\begin{figure}
\includegraphics[clip,width=6.5cm]{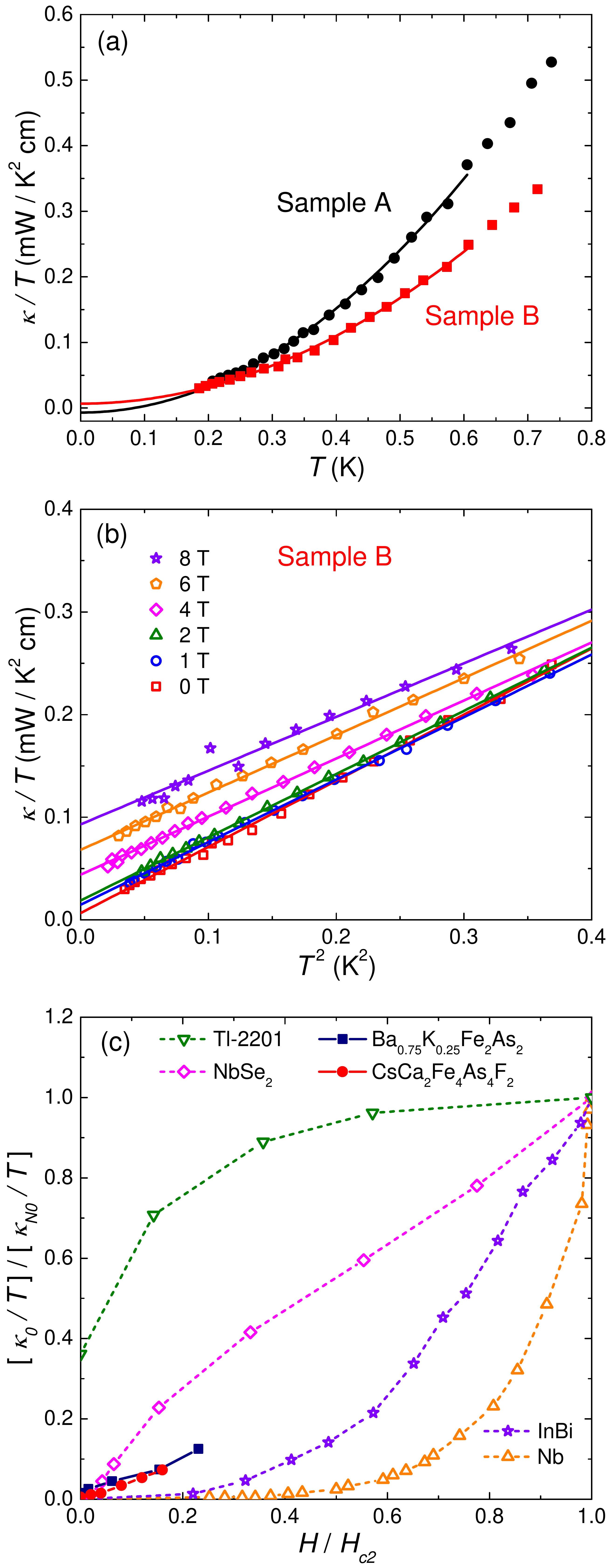}
\caption{(a) Temperature dependence of thermal conductivity for two CsCa$_2$Fe$_4$As$_4$F$_2$ single-crystalline samples (the same to that for resistivity measurements) in zero field. The solid curves represent a fit to $\kappa/T$ = $a$ + $bT^2$ below 0.6 K, giving the residual terms $(\kappa_0/T)_A$ = $-$7 $\pm$ 2 $\mu$W K$^{-2}$ cm$^{-1}$ and $(\kappa_0/T)_B$ = 7 $\pm$ 1 $\mu$W K$^{-2}$ cm$^{-1}$, respectively. (b) Low-temperature in-plane thermal conductivity of sample B in zero and magnetic fields applied along the c axis, plotted as $\kappa/T$ vs. $T^2$. The residual terms given by linear fit below 0.6 K are reproduced in (c). (c) Normalized residual linear term $\kappa_0/T$ in (b) as a function of $H/H_{c2}$. For comparison, similar data are shown for a clean $s$-wave superconductor Nb \cite{Nbkappa}, a dirty $s$-wave superconducting alloy InBi \cite{InBikappa}, a multiband $s$-wave superconductor NbSe$_2$ \cite{NbSe2kappa}, an overdoped $d$-wave cuprate superconductor Tl-2201 \cite{Tl2201kappa}, and a moderately doped iron-arsenide superconductor Ba$_{0.75}$K$_{0.25}$Fe$_2$As$_2$ \cite{luoBaKFeAs}.}
\end{figure}

The ultralow-temperature thermal conductivity measurement is a bulk technique to probe the superconducting gap structure \cite{thermalreview}. Figure 4(a) presents the in-plane thermal conductivity of the two CsCa$_2$Fe$_4$As$_4$F$_2$ single crystals in zero field, plotted as $\kappa/T$ vs $T$. The thermal conductivity at very low temperature usually can be fitted to $\kappa/T$ = $a$ + $bT^{\alpha-1}$, where the two terms $aT$ and $bT^{\alpha}$ represent contributions from electrons and phonons, respectively. For both samples, the fitting parameter $\alpha$ in zero and magnetic fields is very close to 3 (e.g. in zero field $\alpha =$ 3.07 $\pm$ 0.05 for sample A and 3.06 $\pm$ 0.04 for sample B), so we fix it to 3. In order to obtain the residual linear term $\kappa_0/T$ contributed by electrons, we extrapolate $\kappa/T$ to $T$ = 0.

In Fig. 4(a), the fit to $\kappa/T$ = $a$ + $bT^2$ below 0.6 K gives the residual linear terms $(\kappa_0/T)_A$ = $-$7 $\pm$ 2 $\mu$W K$^{-2}$ cm$^{-1}$ and $(\kappa_0/T)_B$ = 7 $\pm$ 1 $\mu$W K$^{-2}$ cm$^{-1}$ for sample A and B, respectively. The Wiedemann-Franz law can tell us the normal-state expectation $\kappa_{N0}/T$ = $L_0$/$\rho_0$, where the Lorenz number $L_0$ is 2.45 $\times$ 10$^{-8}$ W $\Omega$ K$^{-2}$. Here we take sample B for example and obtain $\kappa_{N0}/T$ = 1.28 mW K$^{-2}$ cm$^{-1}$ with $\rho_0$ = 19.2 $\mu\Omega$ cm. Considering this large value of $\kappa_{N0}/T$ and our experimental uncertainty $\pm$5 $\mu$W K$^{-2}$ cm$^{-1}$, the ($\kappa_0/T$)$_A$ and ($\kappa_0/T$)$_B$ in zero field are negligible. Generally, since all electrons form Cooper pairs and no fermionic quasiparticles conduct heat as $T$ $\rightarrow$ 0, there is no residual linear term $\kappa_0/T$ in zero field for nodeless superconductors, as seen in the conventional $s$-wave superconductors Nb and InBi \cite{Nbkappa,InBikappa}. However, for nodal superconductors, the nodal quasiparticles will still contribute a finite $\kappa_0/T$ in zero field. For example, the $\kappa_0/T$ in zero field of the overdoped $d$-wave cuprate superconductor Tl$_2$Ba$_2$CuO$_{6+\delta}$ (Tl-2201) is 1.41 mW K$^{-2}$ cm$^{-1}$ \cite{Tl2201kappa}, and $\kappa_0/T$ is 17 mW K$^{-2}$ cm$^{-1}$ in zero field for the $p$-wave superconductor Sr$_2$RuO$_4$ \cite{Sr2RuO4kappa}. Therefore, the negligible $\kappa_0/T$ of CsCa$_2$Fe$_4$As$_4$F$_2$ single crystal in zero field gives a strong evidence for a fully-gapped superconducting state.

The field dependence of $\kappa_0/T$ can provide more information on the superconducting gap structure \cite{thermalreview}. The thermal conductivity in magnetic fields for sample B is shown in Fig. 4(b), which is also fitted to $\kappa/T$ = $a$ + $bT^2$ below 0.6 K. The $\kappa_0(H)/T$ obtained in Fig. 4(b) is normalized to $\kappa_{N0}/T$, and plotted as a function of $H/H_{c2}$ in Fig. 4(c) with the similar data of a clean $s$-wave superconductor Nb \cite{Nbkappa}, a dirty $s$-wave superconducting alloy InBi \cite{InBikappa}, a multiband $s$-wave superconductor NbSe$_2$ \cite{NbSe2kappa}, an overdoped $d$-wave cuprate superconductor Tl-2201 \cite{Tl2201kappa}, and a moderately doped Ba$_{0.75}$K$_{0.25}$Fe$_2$As$_2$ \cite{luoBaKFeAs}.

In Fig. 4(c), $\kappa_0/T$ of the typical $d$-wave superconductor Tl-2201 starts with a finite value and shows a roughly $\sqrt{H}$ in low fields due to the Volovik effect \cite{Tl2201kappa}. By contrast, in single-gap $s$-wave superconductors like clean Nb and dirty InBi, $\kappa_0/T$ is zero at $H$ = 0 and rises rather slowly in low fields as it relies on the tunneling of quasiparticles between localized states inside adjacent vortex cores. Field dependence of $\kappa_0/T$ of multigap $s$-wave superconductors like NbSe$_2$ and Ba$_{0.75}$K$_{0.25}$Fe$_2$As$_2$ falls in between. A negligible $\kappa_0/T$ at zero field and a rapid rise in low fields can be attributed to the fast suppression of the smaller gap by the applied field \cite{NbSe2kappa}. Because of the lack of the thermal conductivity data of 1144 superconductors, we take the moderately doped 122 superconductor Ba$_{0.75}$K$_{0.25}$Fe$_2$As$_2$ for comparison. From Fig. 4(c), one can see that the normalized $\kappa_0(H)/T$ of Ba$_{0.75}$K$_{0.25}$Fe$_2$As$_2$ and CsCa$_2$Fe$_4$As$_4$F$_2$ manifests very similar behavior, which suggests a multigap nodeless superconducting state in CsCa$_2$Fe$_4$As$_4$F$_2$.

For multigap superconductors, the field dependence of $\kappa_0/T$ depends on the ratio between the large and small gaps \cite{NbSe2kappa}, therefore we examine the magnitudes of the gaps for CsCa$_2$Fe$_4$As$_4$F$_2$ and Ba$_{0.75}$K$_{0.25}$Fe$_2$As$_2$. According to the charge homogenization, the 12442 and 1144 compounds should have similar doping level and Fermi surface topology to moderately doped Ba$_{1-x}$K$_x$Fe$_2$As$_2$ \cite{12442K,1144properties}. Previously, ARPES measurements on CaKFe$_4$As$_4$ single crystals observed three hole pockets ($\alpha$, $\beta$, and $\gamma$) and one electron pocket ($\delta$) \cite{1144arpes}. The superconducting gaps are nearly isotropic on each Fermi surface sheet, but have different magnitudes. The larger gaps of 13 and 12 meV were obtained for the $\beta$ hole and $\delta$ electron sheets, while the $\alpha$ and $\gamma$ hole sheets have smaller magnitudes of 10.5 and 8 meV, respectively \cite{1144arpes}. The ratio between the large and small gaps is roughly 1.6. For moderately doped Ba$_{1-x}$K$_x$Fe$_2$As$_2$, taken $x$ = 0.4 for example, a large superconducting gap ($\sim$ 12 meV) on one hole-like ($\alpha$) and one electron-like ($\gamma$) Fermi surface sheet, and a small gap ($\sim$ 6 meV) on another hole-like sheet ($\beta$) \cite{gaparpes1}. The ratio between the large and small gaps is about 2. If CsCa$_2$Fe$_4$As$_4$F$_2$ has similar superconducting gap structure and magnitudes to CaKFe$_4$As$_4$, above comparable gap ratio well explains the similar field dependence of $\kappa_0/T$ between Ba$_{0.75}$K$_{0.25}$Fe$_2$As$_2$ and CsCa$_2$Fe$_4$As$_4$F$_2$. We note that a more recent $\mu$SR measurements on RbCa$_2$Fe$_4$As$_4$F$_2$ polycrystal suggested that an $s+s$-wave model explains better the temperature dependence of the superfluid density than an $s+d$-wave model \cite{musr12442Rb}. However, their ratio between the large and small gaps (8.15/0.88 = 9.3) is much bigger than that of CaKFe$_4$As$_4$, which can not explain the field dependence of $\kappa_0/T$ for our CsCa$_2$Fe$_4$As$_4$F$_2$ single crystal.

In summary, we measure the ultralow-temperature thermal conductivity of CsCa$_2$Fe$_4$As$_4$F$_2$ single crystals to investigate its superconducting gap structure. A negligible residual linear term $\kappa_0/T$ in zero field and the field dependence of $\kappa_0/T$ mimic those of Ba$_{0.75}$K$_{0.25}$Fe$_2$As$_2$, suggesting multigap nodeless superconductivity in CsCa$_2$Fe$_4$As$_4$F$_2$. These results demonstrate that 12442 compounds, just as CaKFe$_4$As$_4$, should have similar doping level and the superconducting gap structure to moderately doped Ba$_{1-x}$K$_x$Fe$_2$As$_2$, according to the charge homogenization.

We thank M. X. Wang and J. Zhang for helpful discussions. This work is supported by the Ministry of Science and Technology of China (Grant No: 2016YFA0300503 and 2015CB921401), the Natural Science Foundation of China (Grant No. 11421404), the NSAF (Grant No: U1630248).\\

\noindent $^*$ E-mail: shiyan$\_$li$@$fudan.edu.cn

\end{document}